\documentclass[twocolumn,amsmath,amssymb,floatfix,prl,aps,superscriptaddress]{revtex4}
\usepackage{graphicx}
\usepackage{epsfig}
\usepackage{dcolumn}
\usepackage{bm}
\usepackage{times}
\usepackage{color}
\usepackage{soul}
\graphicspath{{C:/Users/Gabriel/Desktop/figures/}}
\begin{document}
\title{Effect of radiation damping on the Child-Langmuir law in open diodes}
\author{Gabriel Gonz\'alez}\email{gabriel.gonzalez@uaslp.mx}
\affiliation{C\'atedras CONACYT, Universidad Aut\'onoma de San Luis Potos\'i, San Luis Potos\'i, 78000 MEXICO}
\affiliation{Coordinaci\'on para la Innovaci\'on y la Aplicaci\'on de la Ciencia y la Tecnolog\'ia, Universidad Aut\'onoma de San Luis Potos\'i,San Luis Potos\'i, 78000 MEXICO}
\begin{abstract}
We present a microscopic derivation of the space charge limited current for the motion of non-relativistic charged particles inside a parallel vacuum tube diode taking into account the radiation reaction force. We study the space charged limited current for two different limiting cases. Our results reveal that in the high field regime the space charge current does not follow the Child-Langmuir law, while in the low field regime the space charge current follows the Child-Langmuir law with and effective electrostatic field, i.e. the so called {\it modified} Child-Langmuir law.
\end{abstract}

\maketitle

\section{Introduction}
\label{sec1}
\begin{figure}[b]
\centering
\includegraphics[width=\linewidth]{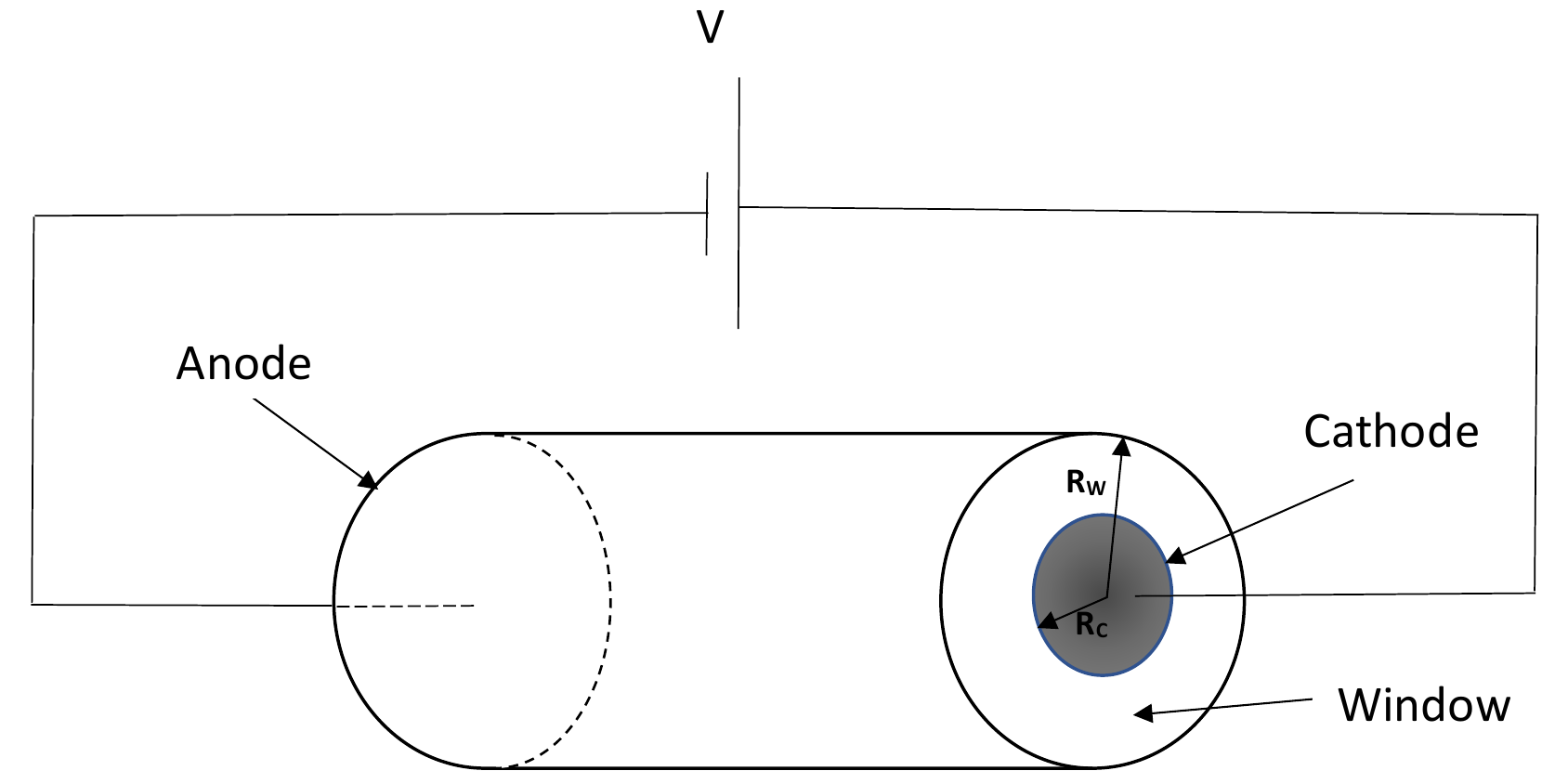} 
\caption{The figure shows the schematic diagram of the open diode. The dark region with radius R$_C$ corresponds to the cathode plate where electrons are emitted from and the white area with radius R$_W$-R$_C$ corresponds to the open window from which radiation may leak.}
\label{fig1}
\end{figure}

The Child-Langmuir law is a statement on the maximum steady state current that can be transported in a planar diode when the initial velocity of the electrons at the cathode is zero.
A simple way to obtain the maximum current density allowed in the parallel plate vacuum tube diode is by means of the {\it microscopic} Child-Langmuir law which is given by \cite{gc}
\begin{equation}
\frac{J}{e}p+\frac{\epsilon_0}{2}E^2=C_1
\label{eq01}
\end{equation}
where $J$ is the space charge current, $e$ is the magnitude of the electron´s charge, $p$ is the linear momentum, $\epsilon_0$ is the free space permittivity, $E$ is the electric field and $C_1$ is a constant that depends on the initial conditions. The constant in eq. (\ref{eq01}) can be set to zero for the case when the initial velocity of the electron and the electric field are zero at the cathode. If we multiply eq. (\ref{eq01}) by the volume charge density $\rho=\epsilon_0 dE/dx$, we end up with
\begin{equation}
\frac{J^2m}{e}+\frac{\epsilon_0^2}{6}\frac{dE^3}{dx}=0
\label{eq02} 
\end{equation} 
where we have used the relation $J=\rho p/m$. We can integrate eq. (\ref{eq02}) from zero to an arbritrary position $x$ to obtain
\begin{equation}
\frac{J^2m}{e}x+\frac{\epsilon_0^2}{6}E^3=0
\label{eq03}
\end{equation}
where we have imposed the boundary condition $E(x=0)=0$. Solving for the electric field in eq. (\ref{eq03}) and using the relation $E=-dV/dx$, we can integrate once again 
for electrodes separated by a distance $D$ and with a fixed electric potential $V_0$, to obtain
\begin{equation}
|J_{CL}|=\frac{4\epsilon_0}{9D^2}\sqrt{\frac{2e}{m}}V_0^{3/2}
\label{eq04}
\end{equation}
Equation (\ref{eq04}) is the well known Child-Langmuir law.\cite{child,lang}
Since the derivation of this fundamental law many important and useful variations on the classical Child-Langmuir law have been investigated to account for special geometries,\cite{lang1,lang2,page} relativistic electron energies,\cite{jory} non zero initial electron velocities,\cite{lang3,jaffe,ahmady} quantum mechanical effects,\cite{lau,ang,gg1} nonzero electric field at the cathode surface,\cite{barbour} slow varying charge density,\cite{gg2} and quadratic damping.\cite{gg3}\\
In a closed system, electrons accelerated under the influence of space charge and the applied potential continuously radiate and reabsorb energy. Most often however, diodes are {\it open} due to the presence of either a mesh anode or
insulating dielectrics in the pulse forming line of the cavity
diode or a dielectric window as in a photocathode (See Fig. (\ref{fig1})).\cite{db} The
electrons are thus unable to reabsorb all the emitted radiation
and hence acquire a lower energy on reaching
the anode as compared to the case of a closed diode.
This also leads to a drop in transmitted current due to
the enhanced repulsion between the slowly moving electrons.
Thus, diodes open to electromagnetic radiation
cannot be governed by the above Child-Langmuir
law. It was shown in Ref. \cite{dbr} that in open diodes where electromagnetic power loss occurs, the Child-Langmuir law can be applied by introducing an effective potential, this is the so called {\it modified} Child-Langmuir law.
In this paper we show from first principles that the {\it modified} Child-Langmuir law accounts for the loss of energy due to the radiation reaction force in open diodes only for the low field regime. We show that for the high field regime the space charge limited current has a new scaling given by $J\propto V_0^{4/3}/D^{5/3}$.\\
\section{Child-Langmuir law with radiation}
We analyze the system in which a beam of electrons inside a vacuum tube diode is accelerated by an electric field and experiences a radiation reaction force. The equation of motion which describe this system is given by
\begin{equation}
m\frac{d^2x}{dt^2}=-eE-\Gamma \frac{d^3x}{dt^3}
\label{eq05}
\end{equation}
where $\Gamma=e^2/6\pi\epsilon_0c^3$. The last term in eq. (\ref{eq05}) is referred as the Abraham-Lorentz force or the radiation reaction force. The form of the radiation reaction force can be derived by applying an energy balance argument to the classical Larmor formula for radiated power and it can also be seen as the force which the field produced by the charge exerts on the charge itself.\cite{poll} It is important to keep in mind that eq. (\ref{eq05}) simulates a beam of electrons and therefore we assume that all the charge and mass of a bunch of electrons is concentrated in a single macro-particle, thus the macro-particle has a charge $e=Ne_0$ and a mass $m=Nm_0$, where $e_0$ and $m_0$ are the charge and mass of the electron, respectively, and $N$ represents the number of electrons in the beam.\\
We can immediately write down a first integral of motion of eq.(\ref{eq05}) which is given by
\begin{equation}
\frac{J}{e}p+\frac{\epsilon_0}{2}E^2+\frac{J\Gamma}{me}\frac{dp}{dt}=C_2
\label{eq06}
\end{equation}
where $C_2$ is a constant that depends on the initial conditions. If we consider that $p=p(E)$, then eq.(\ref{eq06}) is written as
\begin{equation}
\frac{J}{e}p+\frac{\epsilon_0}{2}E^2+\frac{J^2\Gamma}{me\epsilon_0}\frac{dp}{dE}=C_2
\label{eq07}
\end{equation}
where we have used the relation $dp/dt=(J/\epsilon_0)dp/dE$ in the last step. The general solution to eq.(\ref{eq07}) is given by
\begin{equation}
p(E)=\alpha e^{-\frac{m\epsilon_0}{J\Gamma}E}+\beta E^2+\gamma E+\delta
\label{eq08}
\end{equation}
where $\beta=-\epsilon_0e/2J$, $\gamma=\Gamma e/m$, $\delta=e(C_2-J^2\Gamma^2/m^2\epsilon_0)/J$ and $\alpha$ is determined using the initial conditions. For the case of zero initial velocity and zero electric field at the cathode we have $\alpha=-\delta$. If we consider that the acceleration at the cathode is also zero, then $C_2=0$ and we have
\begin{equation}
p(E)=\frac{J\Gamma^2e}{m^2\epsilon_0}\left(e^{-\frac{m\epsilon_0}{J\Gamma}E}-1\right)-\frac{\epsilon_0e}{2J}E^2+\frac{\Gamma e}{m}E
\label{eq09}
\end{equation}
Note that if $\Gamma\rightarrow 0$ we recover the {\it microscopic} Child-Langmuir law.\\
Equation (\ref{eq09}) is far too complex to obtain from it the space charge limited current, nevertheless we can use eq. (\ref{eq09}) to investigate the space charge limited current for two special cases. Let us first consider the case when $|m\epsilon_0E/J\Gamma|<<~1$, in this limit $|J|>>~1.42\times10^{12}|E|/N$. To fulfill this condition let us consider a closed diode with plate separation of $0.1$ mm and applied voltage of $250$ kV, the maximum diode current density is given by eq. (\ref{eq04}) which is $J_{CL}=2.9\times10^{10}$ A/m$^2$. This current density gives for a beam radius of $3.5$ cm and assuming constant acceleration the following number of electrons in the beam $N=4.5\times10^{14}$,\cite{dbr1} thus the condition for the current charge density reads $|J|>>0.00312|E|$. The electric field for the example above is $E=V/D=250\times10^{7}$ V/m, which means that $|J|>>7.76\times10^{6}$ A/m$^{2}$, this condition is experimentally accessible in standard open diodes. Therefore, we can then expand the exponential function given in eq.(\ref{eq09}) for the case when $|m\epsilon_0E/J\Gamma|<<1$, which gives 
\begin{equation} 
p(E)=-\frac{e\epsilon_0^2m}{6J^2\Gamma}E^3
\label{eq10}
\end{equation}
multiplying eq.(\ref{eq10}) by the charge density we have
\begin{equation}
J=-\frac{e\epsilon_0^3}{24J^2\Gamma}\frac{dE^4}{dx}
\label{eq11}
\end{equation}
Integrating eq.(\ref{eq11}) from zero to an arbitrary position $x$ we obtain
\begin{equation}
Jx=-\frac{e\epsilon_0^3}{24J^2\Gamma}E^4
\label{eq12}
\end{equation}
Solving eq.(\ref{eq12}) for the electric field and using the relation $E=-dV/dx$, we can integrate once again 
for electrodes separated by a distance $D$ and with a fixed electric potential $V_0$, to finally obtain
\begin{equation}
|J|=\frac{5\epsilon_0}{8D}\sqrt[3]{\frac{5e}{12\Gamma D^2}}V_0^{4/3}
\label{eq13}
\end{equation}
From eq.(\ref{eq13}) it is clear that the scaling of the classical Child-Langmuir law is not valid when we take into account the radiation reaction force. The new scaling for the case when $|m\epsilon_0E/J\Gamma|<<1$ is given by $J\propto V_0^{4/3}/D^{5/3}$. If we use eq.(\ref{eq13}) with the same values of the example given above we find that the space charge limited current is given by $|J|=1.19\times10^{9}$ A/m$^{2}$, which is around $20$ times smaller than the value given by the regular Child-Langmuir law. \\
Now let us consider the case when $|J\Gamma/m\epsilon_0E|<<~1$, in this limit $|J|<<~1.42\times10^{12}|E|/N$. For this regime we can use the same diode geometry of the example given above but we need to change the applied voltage to $0.1$ V. For the closed geometry the Child-Langmuir law predicts a current density of $J_{CL}=7.376$ A/m$^2$, for this current density we have around $2\times10^8$ electrons, therefore the condition for $|J\Gamma/m\epsilon_0E|<<~1$ which reads $|J|<<7\times10^{6}$ A/m$^{2}$ is easily fulfilled. For the case when $|J\Gamma/m\epsilon_0E|<<~1$, it is better to write eq.(\ref{eq06}) in the following form
\begin{equation}
\frac{J}{e}\left(p+\frac{J\Gamma}{m\epsilon_0}\frac{dp}{dE}\right)+\frac{\epsilon_0}{2}E^2=0
\label{eq14}
\end{equation}
the first term inside the parenthesis in eq.(\ref{eq14}) might be written as $p=p(E+J\Gamma/m\epsilon_0)$ for the case $|J\Gamma/m\epsilon_0E|<<1$. If we make the following substitution $E\rightarrow E+J\Gamma/m\epsilon_0$ we can write eq. (\ref{eq14}) in the form
\begin{equation}
\frac{J}{e}p+\frac{\epsilon_0}{2}\left(E-\frac{J\Gamma}{m\epsilon_0}\right)^2=0
\label{eq15}
\end{equation}
Equation (\ref{eq15}) shows that the {\it microscopic} Child-Langmuir law still holds when radiation power loss occurs, we just have to replace the electrostatic field by an effective electric field $E_{eff}=E-\frac{J\Gamma}{m\epsilon_0}$. This result is consistent with the {\it modified} Child-Langmuir law proposed in Ref.\cite{dbr}.\\ 
We can analyze eq. (\ref{eq15}) in more detail if we multiply by $\rho=\epsilon_0dE/dx=\epsilon_0dE_{eff}/dx$, then eq. (\ref{eq15}) becomes
\begin{equation}
\frac{J^2m}{e}=-\frac{\epsilon_0^2}{6}\frac{dE_{eff}^3}{dx}
\label{eq16}
\end{equation}
Integrating eq.(\ref{eq16}) from zero to an arbitrary position $x$ we have
\begin{equation}
\frac{J^2m}{e}x=-\frac{\epsilon_0^2}{6}\left[E_{eff}^3+\left(\frac{J\Gamma}{m\epsilon_0}\right)^3\right]
\label{eq17}
\end{equation}
where we have used the boundary condition $E_{eff}(0)=-J\Gamma/m\epsilon_0$ in the last step. We can neglect the last term in eq. (\ref{eq17}) since $|J\Gamma/m\epsilon_0|<<1$. Integrating once again eq.(\ref{eq17}) from $0$ to $D$ and using the fact that $E_{eff}=-dV_{eff}/dx=-d(V+J\Gamma x/m\epsilon_0)/dx$ we end up with
\begin{equation}
-\left(V_0+\frac{\Gamma JD}{m\epsilon_0}\right)=\frac{3}{4}\left(-\frac{6J^2mD^4}{\epsilon_0^2e}\right)^{1/3}
\label{eq18}
\end{equation}
Raising to the third power eq.(\ref{eq18}) we have
\begin{equation}
V_0^3\left(1+\frac{\Gamma JD}{m\epsilon_0V_0}\right)^3=\frac{3^3}{4^3}\left(\frac{6J^2mD^4}{\epsilon_0^2e}\right)
\label{eq19}
\end{equation}
If we assume that $|J\Gamma D/m\epsilon_0V_0|<<1$ we have
\begin{equation}
V_0^3\left(1+\frac{3\Gamma JD}{m\epsilon_0V_0}\right)=\frac{3^3}{4^3}\left(\frac{6J^2mD^4}{\epsilon_0^2e}\right)
\label{eq20}
\end{equation}
Solving for $J$ in Equation (\ref{eq20}) we have
\begin{equation}
J=\frac{-\frac{3\Gamma D}{m\epsilon_0V_0}+\sqrt{\left(\frac{3\Gamma D}{m\epsilon_0V_0}\right)^2+\frac{3^4mD^4}{2^3\epsilon_0^2eV_0^3}}}{-\frac{3^4mD^4}{2^4\epsilon_0^2eV_0^3}}
\label{eq21}
\end{equation}
Note that eq.(\ref{eq21}) reduces to the classical Child-Langmuir law when $\Gamma\rightarrow 0$. From eq. (\ref{eq21}) we see that the transmitted current decreases due to the radiation reaction. 

\section{Conclusions}
In conclusion we have shown that in the case when $|m\epsilon_0E/J\Gamma|<<~1$ the space charge limited current does not follow the Child-Langmuir law when radiation loss is taken into account and a new scaling given by $J\propto V_0^{4/3}/D^{5/3}$ is proposed. We have also shown from first principles that when $|J\Gamma/m\epsilon_0E|<<1$ the space charge current density loss can be explained by introducing an effective electrostatic field in the {\it microscopic} Child-Langmuir law, thus validating the {\it modified} Child-Langmuir law given in Ref. \cite{dbr}. 
\section{Acknowledgments}
\vspace{-0.8cm}
This work was supported by the program ``C\'atedras CONACYT".

\end{document}